# A high-energy density antiferroelectric made by interfacial electrostatic engineering

Julia A. Mundy[1,2*], Colin A. Heikes[3*], Bastien F. Grosso[4*], Dan Ferenc Segedin[5], Zhe Wang[6], Berit H. Goodge[6,7], Quintin N. Meier[4], Christopher T. Nelson[8], Bhagwati Prasad[1], Lena F. Kourkoutis[6,7], William D. Ratcliff[3], Nicola A. Spaldin[4], Ramamoorthy Ramesh[1,5,9]& Darrell G. Schlom[7,10]

[1]Department of Materials Science and Engineering, University of California, Berkeley, California 94720 USA
[2]Department of Physics, Harvard University, Cambridge, Massachusetts 02138 USA
[3]NIST Center for Neutron Research, National Institute of Standards and Technology, Gaithersburg, Maryland 20878 USA
[4]Department of Materials, ETH Zürich, CH-8093, Zürich, Switzerland
[5]Department of Physics, University of California, Berkeley, California 94720 USA
[6]School of Applied and Engineering Physics, Cornell University, Ithaca, New York 14853, USA
[7]Kavli Institute at Cornell for Nanoscale Science, Ithaca, New York 14853, USA
[8]Materials Science and Technology Division, Oak Ridge National Laboratory, Oak Ridge, Tennessee 37830 USA
[9]Materials Science Division, Lawrence Berkeley National Laboratory, Berkeley, California 94720 USA
[10]Department of Materials Science and Engineering, Cornell University, Ithaca, New York 14853, USA

* These authors contributed equally to this work

Dielectric capacitors hold a tremendous advantage for energy storage due to their fast charge/discharge times and stability in comparison to batteries and supercapacitors. A key limitation to today's dielectric capacitors, however, is the low storage capacity of conventional dielectric materials. To mitigate this issue, antiferroelectric materials have been proposed[1], but relatively few families of antiferroelectric materials have been identified to date[2–5]. Here, we propose a new design strategy for the construction of lead-free antiferroelectric materials using interfacial electrostatic engineering. We begin with a ferroelectric material with one of the highest known bulk polarizations[6], $BiFeO_3$. We show that by confining atomically-precise thin layers of $BiFeO_3$ in a dielectric matrix that we can induce a metastable antiferroelectric structure. Application of an electric field reversibly switches between this new phase and a ferroelectric state; in addition, tuning of the dielectric layer causes coexistence of the ferroelectric and antiferroelectric states. Precise engineering of the structure generates an antiferroelectric phase with energy storage comparable to that of the best lead-based materials. The use of electrostatic confinement provides a new pathway for the design of engineered antiferroelectric materials with large and potentially coupled responses.

Antiferroelectric materials have seen a resurgence of interest due to proposed applications in a number of energy efficient technologies. These technologies exploit the electric-field-triggered phase transformation from the antipolar ground state[7] to an energetically low-lying polar structure[4,8]. Concomitant changes in the unit cell volume, entropy and stored charge can be used for applications in transducers[9], electrocaloric solid-state cooling[10] and high-energy storage capacitors[1]. Although antiferroelectricity was first discovered in the 1950s in $PbZrO_3$ [ref. 2,3,11], most antiferroelectric materials of interest today are alloyed compounds near a morphotropic phase boundary, which reduces the energy barrier[12,5].

At such morphotropic phase boundaries, the competition between distinct, low-lying ground states can lead to colossal responses from external stimuli. Driving a system to a boundary with chemical alloying, however, fundamentally reduces the polarization of the adjacent ferroelectric phase. Here we demonstrate a new design strategy for the construction of antiferroelectrics using interfacial electrostatics that both avoids chemical substitution and, driven by environmental concerns, is in a material that is lead-free[13]. We start with $BiFeO_3$, one of the strongest known ferroelectric materials with a room-temperature polarization of ~90 $\mu C/cm^2$ (ref. [6]). $BiFeO_3$ can adopt other structures when subjected to large compressive[14] or tensile[15] strain or hydrostatic pressure[16] and additional low-energy polymorphs have been identified by density functional theory (DFT) calculations[6,17–19]. We show that interfacial electrostatic boundary conditions imposed on a confined layer[20–22] can stabilize a new antipolar phase[23]. An applied electric field recovers the ferroelectric state: switching field and thus the storage capacity can be tuned with the interfacial layer.

Figure 1a shows the ferroelectric $BiFeO_3$ structure with $R3c$ symmetry as well as several additional computed low energy non-polar structures. While some have been previously proposed[24], we find a previously unidentified antiferroelectric state labelled "*Pnma*-AFE." For in-plane lattice constants constrained to those of *R3c* BiFeO$_3$, this structure lies 30 meV/f.u. (formula unit) above the *R3c* ground state and is the lowest energy non-polar state identified to date, lower in energy than both the LaFeO$_3$-like structure with *Pnma* symmetry[24] and the PbZrO$_3$-like *Pbam* antiferroelectric structure found in $R_x$Bi$_{1-x}$FeO$_3$ (*R* = rare earth)[25–29]. The *Pnma*-AFE phase (calculated lattice vectors *a* = 5.53 Å, *b* = 11.15 Å and *c* = 15.64 Å) is characterized by antipolar "up-up/down-down" displacements of the bismuth ions and a "super-tilting" pattern of the oxygen octahedra that increases the unit cell in each direction (Fig. 1b). The tilt pattern is composed of alternating octahedral rotations of different amplitudes along the *a* and *b* axes, and a pair of identical clockwise octahedral rotations followed by a pair of identical counterclockwise rotations (where the clockwise pair has a different magnitude) along the *c*-axis; to our knowledge such a pattern has only been previously observed in NaNbO$_3$ (ref. 30). We extend Glazer notation[31] to write the tilt pattern as $a^{\alpha\bar{\beta}\gamma\bar{\delta}}a^{\alpha\bar{\beta}\gamma\bar{\delta}}c^{\varepsilon\varepsilon\bar{\phi}\bar{\phi}}$, where we retain the usual '*a*', '*b*', '*c*' notation to indicate each lattice vector, but we replace the '+' and '-' superscripts by Greek superscripts to indicate the amplitude of the rotation (counterclockwise or clockwise if with an overline) of adjacent octahedra (no rotation would be indicated by a '0' superscript).

We predict that the *Pnma*-AFE phase should be stabilized by appropriate choice of electrostatic boundary conditions, imposed through heteroepitaxy. Since the *Pnma*-AFE structure is non-polar, there is no electrostatic energy cost associated with forming it in thin films, with interfaces to either other non-polar dielectric materials or vacuum. In contrast, the ferroelectric polarization of *R3c* BiFeO$_3$ introduces a depolarizing field in such a geometry, unless free charge carriers are available to screen the large polarization discontinuity at the interface. The total energy of the BiFeO$_3$ in the heterostructure is the sum of the electrostatic energy (higher for the *R*3*c* structure) and the internal energy (lower for the *R*3*c* structure). Since the screening of the depolarizing fields is proportional to the interfacial area, whereas the internal energy is proportional to the BiFeO$_3$ volume, there is a thickness-dependent crossover between the stabilities of the two structures, with the *Pnma*-AFE phase becoming lower in energy in the thin-film limit (Fig. 1c, See Supplement for full discussion).

We demonstrate experimentally that appropriate heterostructures indeed stabilize the new antiferroelectric phase. Superlattices of (La$_x$Bi$_{1-x}$FeO$_3$)$_n$/(BiFeO$_3$)$_n$ for $n = 2 - 20$ were synthesized on (001)$_{pc}$ pseudocubic (pc) TbScO$_3$ substrates using reactive molecular-beam epitaxy. Figure 2a shows simultaneously acquired high-angle annular dark field scanning transmission electron microscopy (HAADF-STEM) and electron energy loss spectroscopy (EELS)[32] images of a (La$_{0.4}$Bi$_{0.6}$FeO$_3$)$_{15}$/(BiFeO$_3$)$_{15}$ film. Chemical intermixing of lanthanum is constrained to within a unit cell of the interface. HAADF-STEM images of the adjacent La$_{0.4}$Bi$_{0.6}$FeO$_3$ layer show a paraelectric structure, consistent with bulk crystals of this composition[28].

High-resolution HAADF-STEM images of the confined BiFeO$_3$ layers viewed along different projection directions show several picometre-scale distortions from the paraelectric phase, which can be assigned to the *Pnma*-AFE phase as shown in Fig. 2b-d. Corresponding mean images obtained by averaging ten equivalent regions are shown in Fig. 2e-g, respectively. In Fig. 2b,e the bismuth atoms are displaced in an "up-up/down-down" pattern, corresponding to the [001] projection of the *Pnma*-AFE structure. This deformation results in peak splitting in the x-ray diffraction scan corresponding to the two distinct out-of-plane lattice constants.

Figure 2c shows an additional picometre-scale modulation observed in other regions, a twin variant of the *Pnma*-AFE structure. Here we observe dumbbells along the bismuth columns with alternating layers of horizontal and diagonal pairs. Figure 2d shows a region similar to that in Fig. 2c, rotated in-plane to image along the substrate $[110]_{pc}$ zone axis. The diagonal dumbbells correlate to a vertically oriented "up-up/down-down" displacement of successive bismuth atoms on alternate planes. Finally, neutron scattering of the superlattice suggests a *G*-type antiferromagnetic structure in the $BiFeO_3$ layers, also found in our DFT calculations.

The relative stability of the *Pnma*-AFE phase and the *R*3*c* phase can be tuned by adjusting the dielectric properties of the adjoining layers (or the $BiFeO_3$ layer thickness). Figure 3a shows a dark-field TEM image of a $(La_{0.4}Bi_{0.6}FeO_3)_{15}/(BiFeO_3)_{15}$ sample consisting of solely the *Pnma*-AFE phase from the same region displayed in Fig. 2b, e. Figure 3b-d shows an $(La_{0.3}Bi_{0.7}FeO_3)_{15}/(BiFeO_3)_{15}$ sample where the more polarizable dielectric layer generates phase coexistence of the *R*3*c* ferroelectric phase and *Pnma*-AFE phase. Here we see areas of the *R*3*c* phase with ferroelectric domains containing 109° and 180° domain boundaries[33] along with regions of the *Pnma*-AFE phase in both of the orientations displayed in Fig. 2b, e and Fig. 2c, f. As shown in the piezoresponse force microscopy image in Fig. 3e, there is phase coexistence at the micrometre-scale. We also find that the $BiFeO_3$ fully adopts the polar *R*3*c* structure in a $(BiFeO_3)_{15}/(SrTiO_3)_{15}$ superlattice given the even more polarizable $SrTiO_3$ layers.

Finally, we demonstrate the conversion of *Pnma*-AFE to a ferroelectric by applying an external electric field, $E_{ext}$. An applied field modifies the system energy by an amount $-\vec{P} \cdot \vec{E_{ext}}$, making the *R*3*c* phase more energetically stable than *Pnma*-AFE for a sufficiently high field as shown in Fig. 4a. We further compute the energy pathway between the polymorphs and find a 26 meV/f.u. kinetic switching barrier (Fig. 4b) via a pathway in which the distortion modes consisting of antipolar movements of the bismuth atoms are suppressed in favour of the cooperative bismuth displacements characteristic of the *R*3*c* phase.

Polarization-electric field hysteresis loops for three $(La_xBi_{1-x}FeO_3)_{15}/(BiFeO_3)_{15}$ samples of varying lanthanum concentration are shown in Fig. 4c-e and Extended Data Fig. 9. At low bias, the $(La_{0.5}Bi_{0.5}FeO_3)_{15}/(BiFeO_3)_{15}$ superlattice in Fig. 4c displays the pinched double hysteresis loop characteristic of the antiferroelectric state[2]. The $(La_{0.3}Bi_{0.7}FeO_3)_{15}/(BiFeO_3)_{15}$ superlattice displays ferroelectric/antiferroelectric phase-coexistence[20]. There is a robust polarization at zero field, but two switching events are observed in the current-voltage response: a higher field switch from the antipolar state and a lower field switch from the ferroelectric state compared to Fig. 4c. Figure 4e shows a ferroelectric $(La_{0.2}Bi_{0.7}FeO_3)_{15}/(BiFeO_3)_{15}$ superlattice, displaying ferroelectric switching at the same voltage as the second switch in Fig. 4d.

We compute the stored energy from the polarization-electric field hysteresis loops. The sample in Fig. 4c has an electrical breakdown voltage of ~2.7 MV/cm with a stored energy density of ~30 J/cm$^3$. These values place our *Pnma*-AFE phase among the best reported perovskite antiferroelectrics, similar to the best relaxor materials and lead-based antiferroelectrics[34]. Given the tunability of the switching voltage with the composition of the adjacent dielectric layer and independently high polarization of the ferroelectric parent phase, this new *Pnma*-AFE structure of $BiFeO_3$ should be a promising candidate for high-energy density capacitor applications.

In summary, we demonstrate the use of electrostatic boundary conditions to uncover a previously metastable state of $BiFeO_3$ that displays high energy storage. Analogous electrostatic engineering could be used to uncover "hidden" ground states in other ferroic oxides with functional structural, electrical or magnetic properties. In contrast to other methods commonly used to manipulate the ground state of oxide materials including isovalent substitution (chemical pressure) or strain engineering, interfacial electrostatic engineering could achieve continuous tuning of the phase stability/coexistence without disorder introduced through inhomogenous dopants or dislocations. Moreover, an applied electric field can return to the bulk ground state – as shown in Fig. 4 – which could lead to large and coupled responses. In the case of our new antiferroelectric/antiferromagnetic $BiFeO_3$, an applied electric field could potentially turn on and off magnetism with the conversion to the ferroelectric/weak ferromagnetic $R3c$ $BiFeO_3$ parent phase.

**Acknowledgements**: We acknowledge discussions with Colin Ophus and Jim Ciston and assistance in the ion bombardment from Sahar Saremi. Funding was primarily provided by the Army Research Office under grant W911NF-16-1-0315. B.F.G., Q.N.M. and N.A.S. acknowledge financial support from ETH Zurich and the Koerber foundation. B.H.G. and L.F.K. acknowledge support by the Department of Defense, Air Force Office of Scientific Research under Award FA9550-16-1-0305. Substrate preparation was performed in part at the Cornell NanoScale Facility, a member of the National Nanotechnology Coordinated Infrastructure (NNCI), which is supported by the National Science Foundation (Grant ECCS-1542081). The electron microscopy imaging studies were performed at the Molecular Foundry, supported by the Office of Science, Office of Basic Energy Sciences, of the U.S. Department of Energy under Contract No. DE-AC02-05CH11231. The electron spectroscopy studies were performed at the Cornell Center for Materials Research, a National Science Foundation (NSF) Materials Research Science and Engineering Centers program (DMR-1719875). The Cornell FEI Titan Themis 300 was acquired through NSF-MRI-1429155, with additional support from Cornell University, the Weill Institute and the Kavli Institute at Cornell. J.A.M. acknowledges the support from a UC President's Postdoctoral Fellowship.

Figure 1.

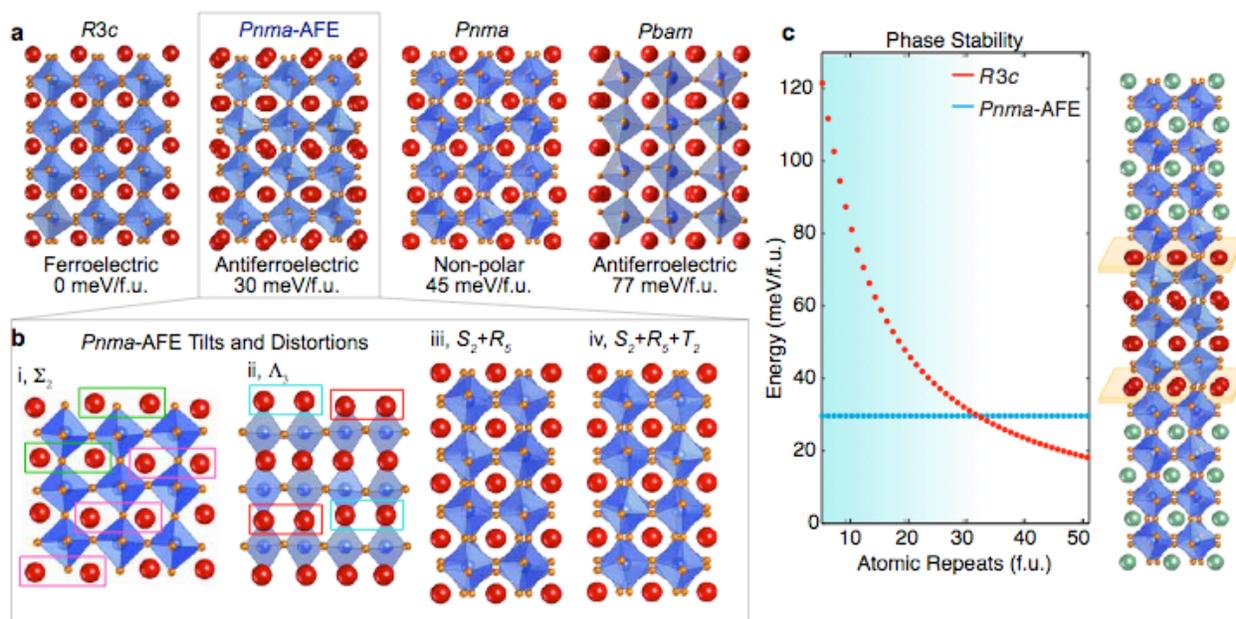

**Figure 1. Energetics of BiFeO$_3$ ground states. a**, Structure and energy of the *R3c* ferroelectric ground state, antiferroelectric "*Pnma*-AFE" and *Pbam* states and non-polar *Pnma* state. Energies are given for structures with the in-plane lattice constants of the *R3c* phase. **b**, Decomposition of the structure of the *Pnma*-AFE phase into its distortions (of symmetries Σ, Λ, R, S and T) relative to the ideal cubic $Pm\bar{3}m$ perovskite: i) up-up/down-down displacements of the atoms atoms along [010]$_{pc}$ (Σ); ii) successive up-up/down-down, zero and up-up/down-down displacements of layers bismuth atoms along the [001]$_{pc}$ direction (Λ), iii) double anti-phase tilting along [001]$_{pc}$, in which two octahedra tilt in one direction with one amplitude and the next pair tilt in the opposite direction with a different rotation angle from the first pair (combination of R and T); and iv) commensurate super-tilting along [100]$_{pc}$ and [010]$_{pc}$, with wavevector π/4*a*, in which each octahedron has a distinct rotation angle and consecutive octahedra are out of phase (combination of S and T). **c**, Phase stability when thin films of BiFeO$_3$ are confined between dielectric layers as shown. For thin films with an appropriate dielectric layer, the *Pnma*-AFE polymorph is lower in energy than the ferroelectric *R3c* polymorph. Bismuth, iron, lanthanum and oxygen are shown in red, blue, green and grey, respectively.

Figure 2.

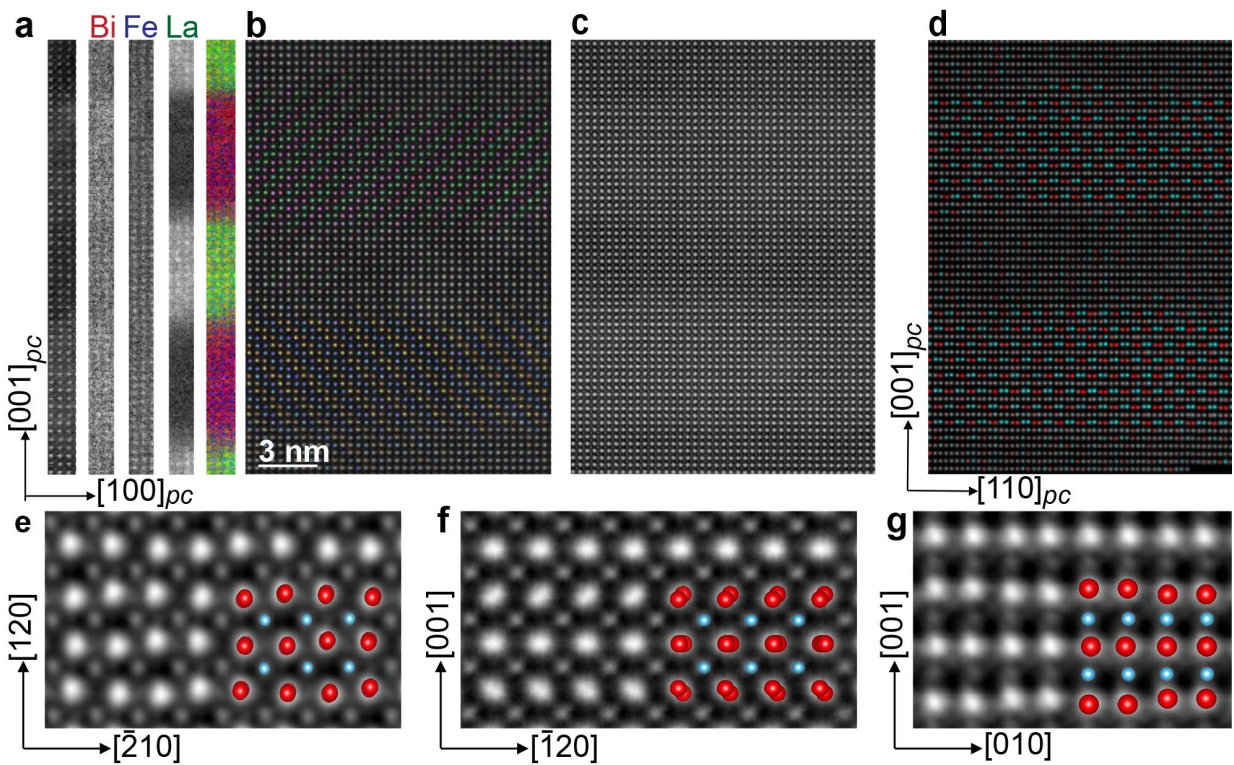

**Figure 2 Atomic-scale characterization of the *Pnma*-AFE phase of BiFeO$_3$ in confined (La$_{0.4}$Bi$_{0.6}$FeO$_3$)$_n$/(BiFeO$_3$)$_n$ superlattices. a**, EELS spectroscopic imaging showing the atomic concentrations of bismuth, iron and lanthanum in red, blue and green, respectively. Two different atomic-scale deformations of the BiFeO$_3$ layer are observed as twin variants in **b** and **c** as viewed along the [010]$_{pc}$ zone axis of the substrate. The atomic-scale displacements in **b** are calculated showing an "up-up/down-down" picoscale distortion with a 45° axis. **d**, A similar region to that shown in **c** viewed along the [110]$_{pc}$ zone axis. Here alternate layers of the bismuth atoms in the BiFeO$_3$ layer show an "up-up/down-down" deformation oriented along the vertical axis. **e-g**, Averaged images from the BiFeO$_3$ layers shown in **b-d**, respectively. The corresponding orientation of the *Pnma*-AFE unit cell is shown on each image.

Figure 3.

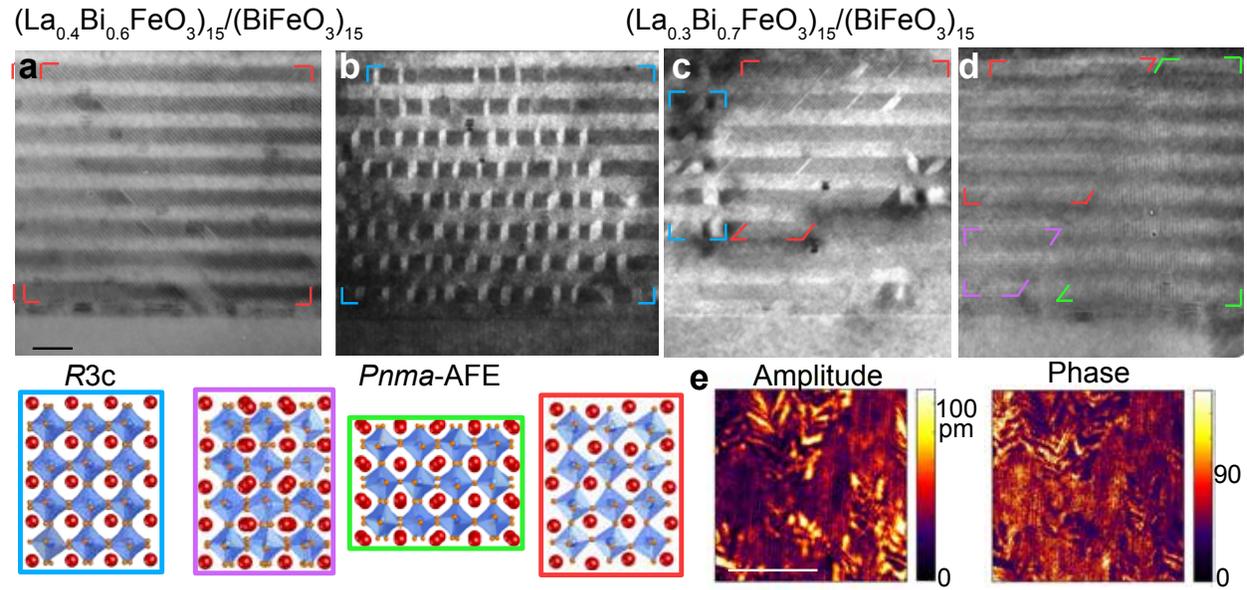

**Figure 3. DF-TEM images of (La$_x$Bi$_{1-x}$FeO$_3$)$_{15}$/(BiFeO$_3$)$_{15}$ superlattices demonstrating structural tunability with altered properties of the surrounding dielectric layer. a**, The (La$_{0.4}$Bi$_{0.6}$FeO$_3$)$_{15}$/(BiFeO$_3$)$_{15}$ superlattice imaged in Fig. 2b,e showing a coherent region of the [001]$_{pc}$-oriented *Pnma*-AFE polymorph of BiFeO$_3$. **b-d**, A (La$_{0.3}$Bi$_{0.7}$FeO$_3$)$_{15}$/(BiFeO$_3$)$_{15}$ superlattice showing phase coexistence between the *Pnma*-AFE and *R*3*c* polymorphs of BiFeO$_3$. As a guide to the eye, regions of the [001]-, [210]- and rotated [210]-oriented *Pnma*-AFE polymorph of BiFeO$_3$ are outlined in red, purple and green, respectively. Regions with 109° and 180° domain walls of the *R*3*c* ferroelectric phase are outlined in blue. Scale bar is 15 nm. **e,** Piezoresponse force microscopy of the sample shown in **b-d** demonstrating regions of high piezo-contrast coexisting with regions of low contrast. Scale bar 0.5 μm.

Figure 4

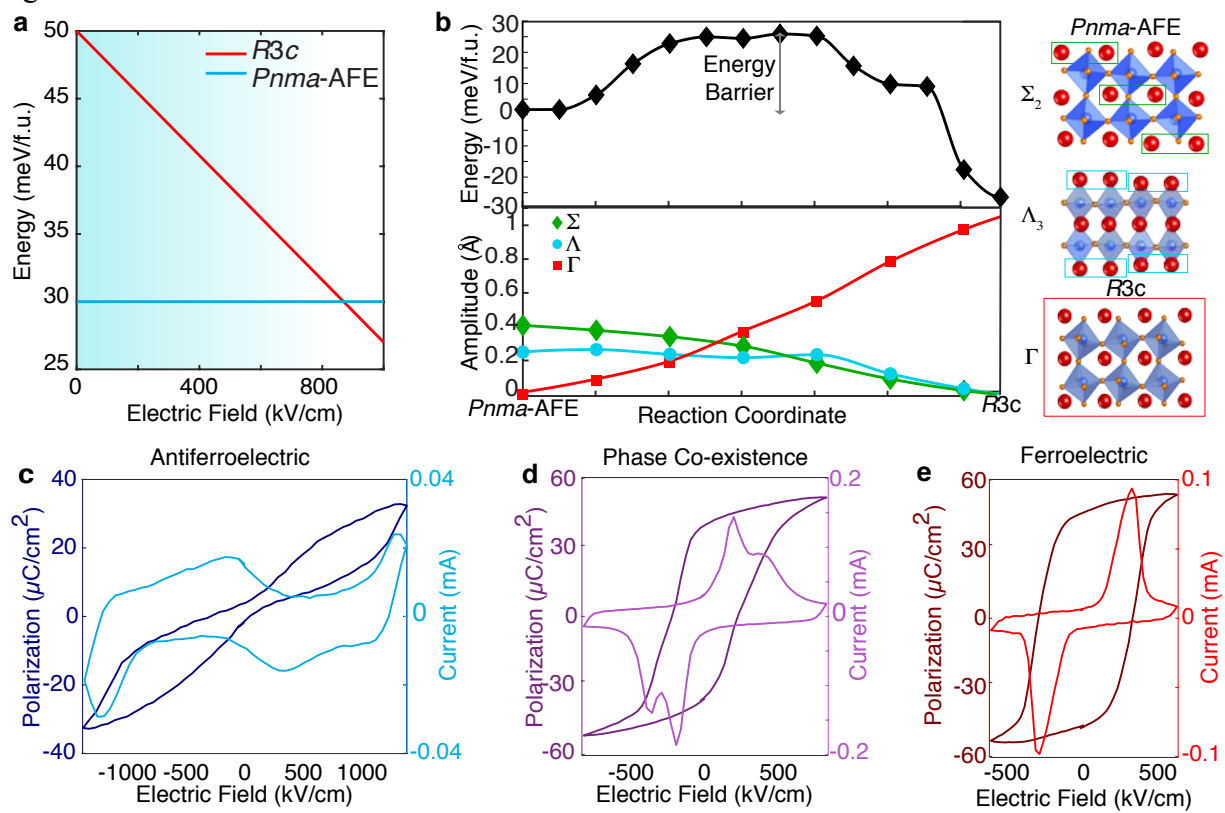

**Figure 4. Electric-field-induced switching between the *Pnma*-AFE state and the ferroelectric state of BiFeO₃.** **a**, Application of an electric field alters the relative stability of the *R3c* and *Pnma*-AFE phases in heterostructures. For a given layer thickness/dielectric, the ferroelectric *R3c* phase can become more stable than the antiferroelectric *Pnma*-AFE phase (plotted for 15 f.u.). **b**, Calculation of the barrier using the nudged elastic band method. There is a 26 meV/f.u. activation barrier to switch between the phases. The reaction pathway is shown, tracing the damping of the Σ and Λ antipolar distortion modes of *Pnma*-AFE and increase of the polar Γ mode of *R3c* structure. **c-e**, Polarization-electric field hysteresis loops for $(La_xBi_{1-x}FeO_3)_{15}/(BiFeO_3)_{15}$ superlattices with $x = 0.5$, $0.3$ and $0.2$, respectively. Tuning the lanthanum concentration in the dielectric layer converts the structure from an antiferroelectric phase as shown in **c-d** to a ferroelectric structure in **e**.

**Supplemental Information: Calculation of the relative stability of polar and non-polar layers in insulating heterostructures**

Here we present the derivation of the crossover in stability from the polar *R3c* structure of BiFeO$_3$, which is the bulk ground state, to the non-polar *Pnma*-AFE structure that becomes stable in thin-film (BiFeO$_3$)$_n$/(La$_x$Bi$_{1-x}$FeO$_3$)$_n$ heterostructures, as a function of BiFeO$_3$ film thickness. This derivation provides the physics behind Figure 1c of the main text.

In a finite slab of ferroelectric material, the termination of the polarization at the surfaces of the slab results in the formation of bound charge on both surfaces. These in turn generate a depolarizing field, given by $\epsilon_d = \frac{-P}{\varepsilon_0 \varepsilon_r}$, where the minus sign indicates that the depolarizing field acts to oppose the polarization. Here $P$ is the component of polarization perpendicular to the interface, $\varepsilon_r$ is the relative permittivity of the material in which the field is created and $\varepsilon_0$ is the permittivity of free space. (In the case of a heterostructure with multiple ferroelectric layers, each of the layers has a depolarizing field associated with it.) There is an electrostatic energy cost per unit area,

$$E_{es} = \int_0^d dz \frac{\varepsilon_0 \varepsilon_r}{2} \epsilon_d^2 = \frac{\varepsilon_0 \varepsilon_r}{2} \epsilon_d^2 \, d \qquad (S.1)$$

associated with the existence of the depolarizing field, which depends on the thickness, $d$, of the ferroelectric layer.

In practice the system adopts a screening mechanism to reduce the electrostatic energy cost associated with the depolarizing field. In the absence of metallic electrodes to screen the bound surface charges, here we consider a mechanism in which free charges are generated by electron-hole pair excitation across the intrinsic band gap, $E_g$, of BiFeO$_3$. We note that this likely provides an upper bound on the screening energy cost, since free carriers will also be available from extrinsic sources such as point defects. The screening field is given by, $\epsilon_{scr} = \dfrac{\sigma}{\varepsilon_0 \varepsilon_r}$, where $\sigma$ is the areal density of free charges generated by electron-hole pair excitation, so that the total electrostatic energy cost per unit area associated with the presence of the screened polarization in the slab is:

$$E_{es} = \frac{1}{2}\varepsilon_0 \varepsilon_r \left(\epsilon_d + \epsilon_{scr}\right)^2 d + E_{scr} = \frac{1}{2}\varepsilon_0 \varepsilon_r \left(\frac{-P}{\varepsilon_0 \varepsilon_r} - \frac{\sigma}{\varepsilon_0 \varepsilon_r}\right)^2 d + E_{scr} \qquad (S.2)$$

Here $E_{scr} = \dfrac{|\sigma|}{e^-} E_g$ is the energy cost per unit area of creating areal charge $\pm\sigma$ simultaneously on each interface.

The amount of screening charge is determined by a balance between the energy cost of maintaining the polarization in the presence of the depolarizing field and the energy cost of creating electron-hole pairs across the gap. To determine this we minimize

$$E_{es} = \frac{P^2}{2\varepsilon_0 \varepsilon_r} d + \frac{\sigma^2}{2\varepsilon_0 \varepsilon_r} d - \frac{P\sigma}{\varepsilon_0 \varepsilon_r} + \frac{\sigma}{e^-} E_g \qquad (S.3)$$

with respect to areal charge density (assuming without loss of generality that $P > 0$ and $\sigma > 0$) by setting $\dfrac{\partial E_{es}}{\partial \sigma} = 0$. We obtain

$$\sigma = \begin{cases} P - \dfrac{\varepsilon_0 \varepsilon_r}{2d} \dfrac{E_g}{e^-} & \text{if } d \geq \dfrac{\varepsilon_0 \varepsilon_r}{P} \dfrac{E_g}{e^-} \\ 0 & \text{if } d < \dfrac{\varepsilon_0 \varepsilon_r}{P} \dfrac{E_g}{e^-} \end{cases} \quad (S.4)$$

for the equilibrium surface density of free screening charges.

Substituting for σ in our expression for the electrostatic energy per unit area we obtain

$$E_{es} = \begin{cases} P\dfrac{E_g}{e^-} - \dfrac{\varepsilon_0 \varepsilon_r}{2d}\left(\dfrac{E_g}{e^-}\right)^2 & \text{if } d \geq \dfrac{\varepsilon_0 \varepsilon_r}{P} \dfrac{E_g}{e^-} \\ \dfrac{1}{2}\dfrac{P^2}{\varepsilon_0 \varepsilon_r} & \text{if } d < \dfrac{\varepsilon_0 \varepsilon_r}{P} \dfrac{E_g}{e^-} \end{cases} \quad (S.5)$$

This is the additional electrostatic energy per unit area associated with the formation of the polar *R3c* phase over the non-polar *Pnma*-AFE phase. The total energies per unit area of each phase are then

$$\begin{aligned} E_{R3c} &= u_{R3c} d + E_{es} \\ E_{AFE} &= u_{AFE} d \end{aligned} \quad (S.6)$$

Where $u_{R3c}$ and $u_{AFE}$ are the volume energy densities of the ferroelectric *R3c* and antiferroeletric *Pnma*-AFE phases respectively, where we assume that the ferroelectric polarization $P$ to be fixed. Taking the DFT energy values of -555.93 meV/Å³ (-33.553 eV/f.u) and -555.38 meV/Å³ (-33.519 eV/f.u.) respectively for the *R3c* and *Pnma*-AFE phases, together with and a band gap energy of $E_g$ = 2 eV, dielectric constant $\varepsilon_r$ = 55 and polarization $P_z$ = 60 µC/cm² yields the result shown in Figure 1c. of the main text. We obtain an energetic crossover at around 130 Å thickness (around 30 formula units of the the *R3c* phase), where the energy density of the polar *R3c* phase drops below that of the *Pnma*-AFE phase.

We note two other mechanisms for reducing the electrostatic energy that could occur in our insulating samples: First, polarization of the inter-layer dielectric region, which is favored by a more polarizable dielectric with large dielectric constant[2,3], and second, the formation of domains within the ferroelectric layer. Both of these effects introduce a field into the inter-layer non- ferroelectric layer.

Figure 4a of the main text is obtained by adding a term $(-\vec{P} \cdot \vec{E_{ext}})$, where $\vec{E_{ext}}$ is an external applied field to the total energy for the case of a 15 formula unit thick film in the conditions plotted in Fig. 1c. The polar phase is favoured for a field of 1.4 kV/cm.